\pdfoutput=1
\documentclass[conference]{IEEEtran}
\IEEEoverridecommandlockouts

\usepackage{cite}
\usepackage{amsmath,amssymb,amsfonts}
\usepackage{algorithmic}
\usepackage{graphicx}
\usepackage{textcomp}
\usepackage{xcolor}
\def\BibTeX{{\rm B\kern-.05em{\sc i\kern-.025em b}\kern-.08em
    T\kern-.1667em\lower.7ex\hbox{E}\kern-.125emX}}

\usepackage{dblfloatfix}
\usepackage{svg}
\usepackage{placeins}
\usepackage{todonotes}
\usepackage{tabularx}
\usepackage{multirow}
\usepackage{footnote}
\usepackage{hyperref}
\usepackage{amssymb}
\usepackage{amsmath}
\hypersetup{
  colorlinks = true,
  linkcolor  = blue,
  urlcolor   = blue,
  citecolor  = black,
}
\usepackage{url}
\usepackage{cleveref}
\newcommand*{\fullref}[1]{\hyperref[{#1}]{\cref*{#1} \nameref*{#1}}}
\newcommand*{\Fullref}[1]{\hyperref[{#1}]{\Cref*{#1} \nameref*{#1}}}
\newcommand*{\secref}[1]{\hyperref[{#1}]{\autoref*{#1}}}
\newcommand*{\Secref}[1]{\hyperref[{#1}]{\Cref*{#1}}}

\let\oldFootnote\footnote
\newcommand\nextToken\relax

\renewcommand\footnote[1]{%
    \oldFootnote{#1}\futurelet\nextToken\isFootnote}

\newcommand\isFootnote{%
    \ifx\footnote\nextToken\textsuperscript{,}\fi}

\begin{document}

\title{SWARM-SLR AIssistant: A Unified Framework for Scalable Systematic Literature Review Automation
\thanks{This work was funded by the DFG SE2A Excellence Cluster.}
}

\author{
    \IEEEauthorblockN{
    Tim Wittenborg\IEEEauthorrefmark{1},
    Allard Oelen\IEEEauthorrefmark{2},
    Manuel Prinz\IEEEauthorrefmark{2}
    \IEEEauthorblockA{\IEEEauthorrefmark{1}L3S Research Center, Leibniz University Hanover, Hanover, Germany
    \\tim.wittenborg@l3s.uni-hannover.de}
    \IEEEauthorblockA{\IEEEauthorrefmark{2}TIB - Leibniz Information Centre for Science and Technology, Hanover, Germany
    \\\{allard.oelen, manuel.prinz\}@tib.eu}
    }
}

\maketitle

\begin{abstract}
Despite a growing ecosystem of tools supporting Systematic Literature Reviews (SLRs), integrating them into user-friendly workflows remains challenging.
The Streamlined Workflow for Automating Machine-Actionable Systematic Literature Reviews (SWARM-SLR) unified the tool annotation and provided a cohesive yet modular workflow, but faced scalability and usability issues. 
We introduce the SWARM-SLR AIssistant, a unified framework that combines the SWARM-SLR's structured methodology with an agent-based assistant that integrates research tools in a modular interface. 
The first SWARM-SLR stage is integrated, enabling conversational, LLM-guided support and persistent data storage.
To address the tool assessment bottleneck, we propose a centralized tool registry that allows developers to annotate and register tools autonomously using a shared metadata schema.
Preliminary evaluation shows improved usability, but challenges remain in balancing efficiency, accessibility, and transparency.
Further development is needed to realize scalable SLR automation.
\end{abstract}


\begin{figure*}[b!]
  \includegraphics[width=\textwidth]{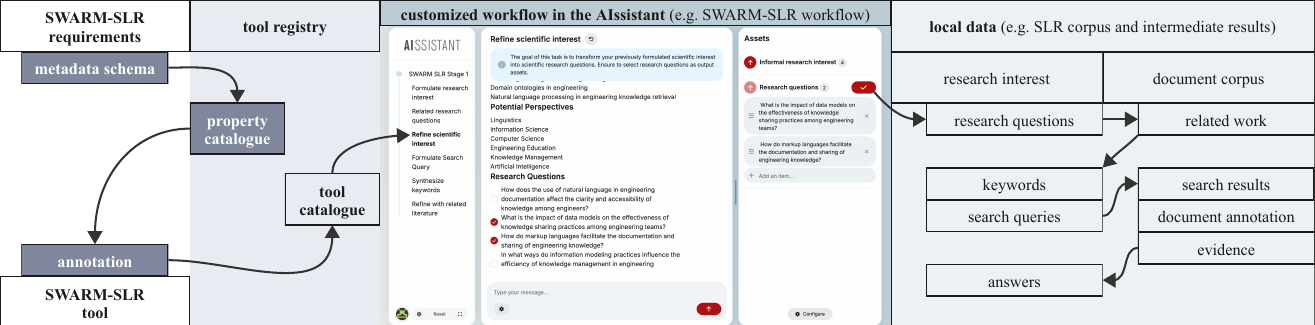}
  \caption{Overview of the SWARM-SLR AIssistant. The SWARM-SLR requirements are mapped to tool registry properties, which are used to annotate tools. The resulting tool catalogue can be used from within the AIssistant, providing a uniform interface for human and machine to access tools and data.}
  \label{fig:teaser}
\end{figure*}

\section{Introduction\label{sec:introduction}}
Despite the increasing availability of research tools tailored to various stages of Systematic Literature Reviews (SLRs), their integration into a coherent workflow remains a major hurdle.
The Streamlined Workflow for Automating Machine-Actionable Systematic Literature Reviews (SWARM-SLR) provided a complete, structured SLR process~\cite{wittenborg_swarm-slr_2024}, and a first complete application of the workflow showed its capabilities~\cite{wittenborg_knowledge-based_2025}.
Yet, the practical adoption of the tool annotation and utilization has been slow due to usability challenges, from setting up the workflow to navigating the Jupyter notebook environment to installing and switching between several tools.

We developed the AI supported research assistant (i.e., AIssistant), which provides LLM-based guidance throughout different tasks and workflows, calling external, configurable tools where needed.
SWARM-SLR and the AIssistant face integration bottlenecks, the AIssistant relying on JSON schema annotations of available tools and the SWARM-SLR on filled out surveys positioning tools capacities over the workflow.
Neither approach scales well into an autonomous solution.

This work proposes a unified tool registry to utilize the synergistic potential of aligning SWARM-SLR and AIssistant tool annotation, providing a unified framework enabling developers to contribute their solutions and assists users to customize and utilize the system.
Our contributions include:
\textbf{1)} The first stage of the SWARM-SLR integration into the AIssistant, providing workflow guidance in a uniform interface with a persistent storage layer that allows human and machine shared access to intermediate and final results.
\textbf{2)} A conceptual design of an AIssistant tool registry, centralizing the tool annotation for both approaches and allowing developers to annotate tools autonomously for future user selection.
\textbf{3)} A preliminary evaluation of the current implementation, with 18 survey respondents providing feedback on the implementation of the first 5 SWARM-SLR steps and potential future work.

This work is structured as follows:
\Secref{sec:background} details the background required for our approach in \secref{sec:approach}.
The evaluation design and results are detailed in \secref{sec:evaluation}.
These findings and future work are discussed in \secref{sec:future}, concluding in \secref{sec:conclusion}.

\section{Background and Related work\label{sec:background}}
\subsubsection{SWARM-SLR\label{sec:swarm}}
The original SWARM-SLR workflow provided a structured methodology to conduct machine-actionable SLRs, identifying 65 requirements that are mapped to 19 steps.
These steps are part of the 8 major tasks making up an SLR, which are categorized into 4 Stages.
This classification allows mapping tools to the requirements and hence formalizing their potential strengths and weaknesses over the SLR stages, tasks and steps.
However, practical uptake suffered due to configuration overhead and decentralized tooling.
Despite SWARM-SLR being presented at several workshops, the tool adding process never picked up.
There were frequent questions about the efficiency of collaboration with others and how to prevent moving between the many different applications.
The AIssistant has the potential to address these challenges by providing an integrated platform where different tools are operated without leaving the environment.

\subsubsection{ORKG Ecosystem\label{sec:orkg}}
The modular SWARM-SLR was designed to utilize the Open Research Knowledge Graph (ORKG)~\cite{auer2020improving}.
Since SWARM-SLRs publication, ORKG ASK~\cite{oelen2024orkg} was released, a natural-language interface to literature search over a predefined corpus of more than 70,000,000 papers.
Similarly, the support for local corpus creation and processing was further developed by John et al.~\cite{john_human---loop_2025}.
Each time a new tool such as ORKG ASK~\cite{oelen2024orkg} is developed, it needs to be integrated into the workflow as well, potentially with new frontends with which users need to familiarize themselves.

\subsubsection{Research Tool Ecosystem\label{sec:extensions}}
The previous SWARM-SLR tool annotation aimed to improve upon previous works from Bosman and Kramer~\cite{bosman_tools_2023,bosman_400_tools_and_innovations_in_scholarly_communication_2023} or lists such as the Systematic Review Toolbox\footnote{\url{https://systematicreviewtools.com/}}.
The latter temporarily ceased to exist right before the SWARM-SLR paper was published, revealing a gaping flaw in any such system:
If the central hub breaks down, the system becomes unusable.
An analysis of similar crowdsourced tool annotation approaches~\cite{wittenborg_research_2025}, from tool or package hosting services like Python Package Index (PyPI)\footnote{\url{https://pypi.org/}} to others just hosting information about them, such as awesome lists\footnote{\url{https://github.com/sindresorhus/awesome}}, revealed similar issues, yet also several solutions:
Designing a standardized annotation format that can be harvested asynchronously and by anyone. 
Instead of centralizing all metadata, annotations can exist alongside the tool itself (e.g., in a GitHub repository), allowing communities to retain control over the information infrastructure.
Especially tool registry approaches have proven effective, such as the Visual Studio Code (VS Code) Marketplace\footnote{\url{https://marketplace.visualstudio.com/vscode}}, the Obsidian Community Plugins\footnote{\url{https://obsidian.md/plugins}} or \href{https://bio.tools/}{bio.tools}~\cite{ison_tools_2016}.
These extension-based environments streamline tool distribution and integration.
Containerized solutions such as Docker\footnote{\url{https://www.docker.com/}} and orchestration frameworks like Kubernetes\footnote{\url{https://kubernetes.io/}} address a different layer of the ecosystem, providing consistent runtime environments and scalable deployment.

\section{Proposed System\label{sec:approach}}
We seek to integrate the SWARM-SLR workflow into the AIssistant and improve both systems' modular tool annotation with a uniform tool registry, as illustrated in \secref{fig:teaser}.

\subsubsection{AIssistant\label{sec:tibai}}
The AIssistant provides a workspace for human-machine collaboration.
It integrates tools into workflows via a modular interface, storing intermediate and final assets, such as text or documents. 
While several agents are predefined in the default setup, integrating services such as ORKG ASK, Semantic Scholar, and arXiv queries, is one of the core design aspects is the modular tool customization.
Tools are defined in modular files, describing their functionalities and parameters textually, in addition to a tool execution functionality.
The textual description of the function interfaces ensures its compatibility with LLM tool callings, allowing the LLM to invoke tools when deemed necessary.
The function calling process varies among tools, but generally includes calling specific REST endpoints to gather information, often based on user specified input.
Tool executions are kept simple to decouple systems, ensuring that the tool executing, and in turn, tool integration, is easy. 

\subsubsection{SWARM-SLR AIsssitant\label{sec:slrai}}
The SWARM-SLR steps are integrated as separate agents in the interface.
Each has it own system prompts, list of enabled tools, and list of input/output assets.
Separating the different steps into different agents has as benefit that the user is more in control, and that it is easier to modify intermediate data.
In the end, our goal is not to provide a fully automated approach, but instead provide AI-supported guidance to users in the SWARM-SLR process. 

In its current layout, this integration requires forking the open-source code, then a merge request, requiring review and confirmation from the maintainers.
This approach still warrants improvements, as discussed in the \nameref{sec:infrastructure} section, but is sufficient for an early SWARM-SLR integration test.

As shown in \secref{fig:teaser}, we integrated the first stage of the SWARM-SLR workflow into the AIssistant, using the manual JSON configuration for the first 5 steps.
Steps such as "Formulate Search Query" may be split into multiple steps (see "Synthesize keywords"), if their separation reduces confusion for both human and machine due to nested task descriptions.
Others, such as "Step 2.5 Re-evaluate with domain experts" concluding stage I, are inherently not automatable, and hence not implemented.
Beyond that, several other stages are already supported by the AIssistant, such as literature search and corpus creation.
Succeeding these steps into the installation of further tools, such as the custom python package\footnote{\url{https://pypi.org/project/bnw-tools/}} or Obsidian, warranted a further rethinking of the workflow.
In the current design, the SWARM-SLR AIssistant is not capable of integrating these tools into the same interface.
To still reduce the barrier to a complete workflow, we redesigned the tool curation.

\subsubsection{Tool Registry}
\label{sec:infrastructure}

\begin{figure}
    \centering
    \includegraphics[width=\linewidth]{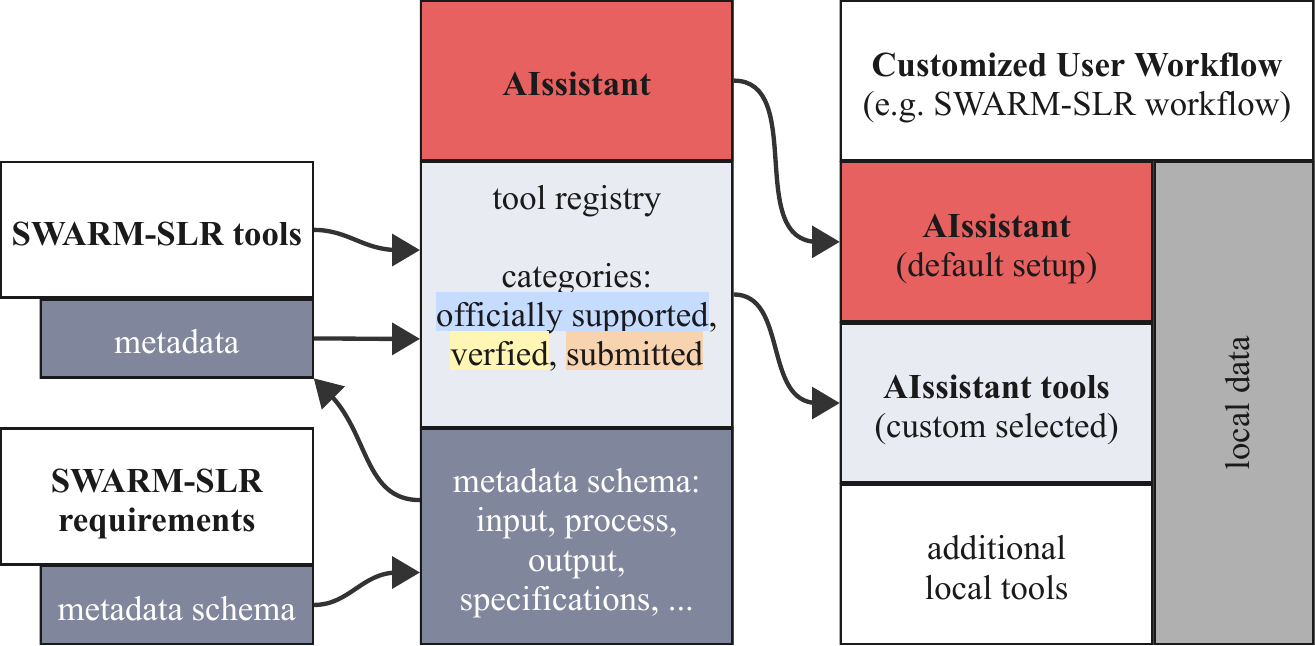}
    \caption{The conceptual design of the AIssistant tool registry provides a central tool curation point, mapping requirements to metadata schema. Users can browse task-specific tools similar to an extension marketplace.}
    \label{fig:tool registry}
\end{figure}
\begin{figure*}[t!]
    \centering
    \includegraphics[width=\linewidth]{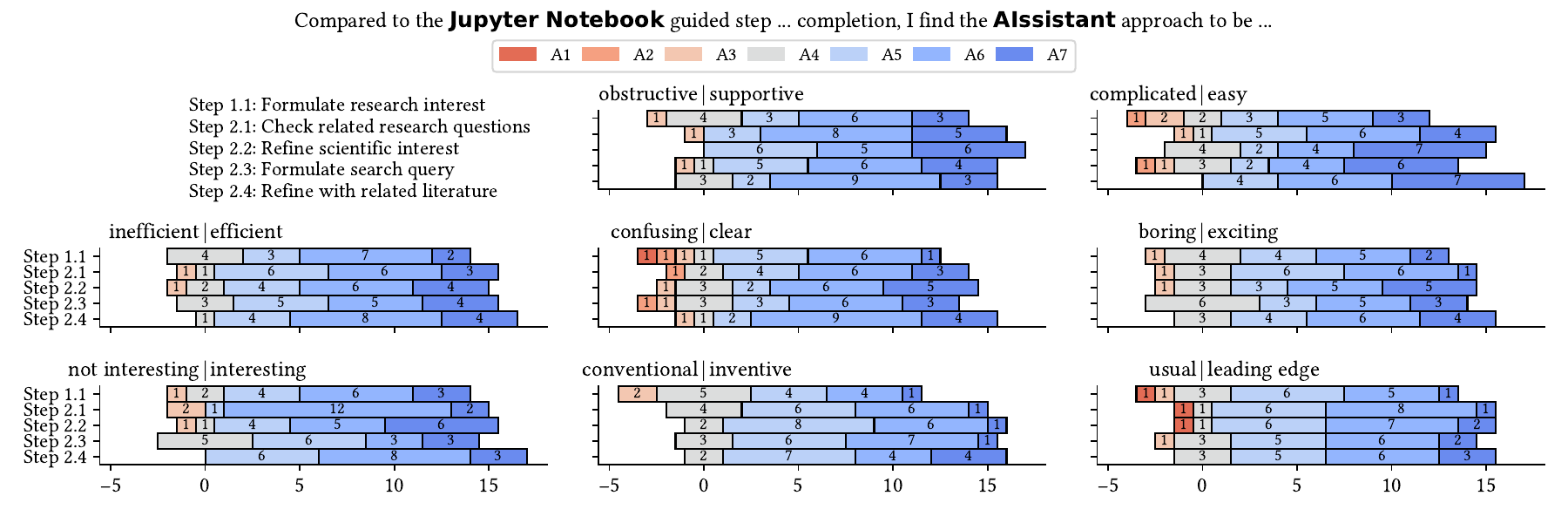}
    \caption{UEQ Results (N=18). The sentiment is noteworthy positive. Commulative over every step, each metric has 2-8 reported negative and 62-74 positive leanings. \textit{Confusing | clear} has the most negative votes (8), with the sole strong negative vote, except for 3 strong \textit{usual} votes. \textit{Obstructive | supportive} has the most positive votes total (74), with 21 of them being strong. \textit{Complicated | easy} has 6 more strong positive votes (27), where \textit{conventional | inventive} has the least (8).}
    \label{fig:ueq_results}
\end{figure*}
Our concept displayed in \secref{fig:tool registry} is built on the example of \href{https://bio.tools/}{bio.tools}~\cite{ison_tools_2016}, which successfully facilitated the crowdsourcing of over 30,000 tools in the past decade.
The conceptual design is a centralized tool registry, asynchronously populated with data autonomously provided by users.
Tool developers can annotate their tools by providing a uniformly structured metadata, for example, using the SWARM-SLR requirement schema. 
This allows users to provide and access tool annotation data via a central system, while at the same time not putting their work at risk in case of the central system failure.
Synergistic effects can be utilized by merging the SWARM-SLR tool annotation into the larger ecosystem of the AIssistant, allowing the SWARM-systematic literature review to be one of several layouts the AIssistant may support.
These changes simplify tool onboarding and lower the technical threshold for contributing tools and workflows.

\section{Evaluation\label{sec:evaluation}}
The current iteration of SWARM-SLR was tested over a year in the Aerospace Engineering SLR use-case.
Since the AIssistant itself is in its early development stages and not released yet, there is no public, large-scale test of it or its SWARM-SLR layout.
The tool integration required for a complete a SWARM-SLR workflow is increasingly complicated with tool capabilities, reaching from simple LLM queries to tools processing, analyzing, restructuring, creating and deleting thousands of documents on the local hard drive.
The required workflow coordination and guard railing of tools so heterogenous promises to be very time consuming.

\subsubsection{Design}
The entire survey design is available online\footnote{\url{http://github.com/borgnetzwerk/tools/tree/main/scripts/SWARM-SLR/data}}
by clicking on \href{https://html-preview.github.io/?url=https://github.com/borgnetzwerk/tools/blob/main/scripts/SWARM-SLR/data/LimeSurvey%20-%20SWARM-SLR%20AIssistant.html}{SWARM-SLR AIssistant Survey (view only)}.
Its pages include data protection, general information, and questions regarding steps 1.1 to 2.4., demographic data, and future work.
The general information explains the survey context, design, and limitation to screenshots.
The following 5 pages depict the SWARM-SLR, first in its original Jupyter Notebook design, then implemented in the AIssistant.
Below these screenshots, two question types were presented, a semantic differential UEQ-S~\cite{schrepp_design_2017} in randomized order and a free text field for any other comment.
All questions were optional, and the "no answer" option was always available.
The demographic data page inquired the highest completed level of education and their prior experience with subjects such as literature reviews.
Future Work explained the proposed tool registry using \secref{fig:teaser} and text, followed by two questions on future motivation to develop and utilize the system.

\subsubsection{Results}
18 participants (10 PhDs, 8 Masters) completed the survey (avg. 22 min, $\sigma$: 14:20 min, median: 18:34 min).
The UEQ-S results are visualized in \secref{fig:ueq_results}.
Evidently, the AIssistant is an improvement upon the previous Jupyter Notebook implementation.
Seven participants used the text fields to provide additional feedback, reaching from UI suggestions to elaborate advice.
One respondent suggested that the Model Context Protocol (MCP) in a local IDE could achieve comparable results "\textit{and still retain all your working environment, extensions, and not needed to know the terminology and "workflows" of the assistant.}"
Since the main goal of the SWARM-SLR AIssistant is to provide more guidance to users, it may be less relevant if an equal result could be reached through a process that is more complicated to set up.
The general sentiment was very positive: "\textit{The AIssistant is a very nice simplification of the Notebook UI, which should make the tool easier to use and therefor more accessible.}" 
Eleven participants would like to use the SWARM-SLR AIssistant, three of which would also use the Jupiter Notebook approach.
Two would like to add their own tools to the workflow, two others their own steps.
At the same time, remarks were made regarding additional energy consumption and other costs, as well as AI-inherent transparency, hallucination, and reproducibility issues.
The evaluation highlights a very valuable perspective:
The further research work is abstracted away, the less oversight humans have over quality and reliability.
At a certain tipping point, the time saved automatically synthesizing information is spent ensuring the process and its resulting knowledge is reliable.
The future of accessible, reliable SLR automation will scale first and foremost with its ability to efficiently communicate its process.

\subsubsection{Threats to Validity}
There are some noteworthy limitations that may affect the generalizability of our work.
The small sample size of 18 participants is inherently limiting, and despite the overwhelmingly positive results, a selection bias in the participants or similar could have influenced this outcome.
Rejecting this possibility requires upscaling of the evaluation, which is inhibited due to the unreleased state of the AIssistant.
Our work presents an evaluation in an early stage of development, aiding similar developments in the rapidly changing field of human-AI collaboration, yet does not equate a full evaluation of our future system.
Evaluating this full system is going to be excessively time-intensive, however, and may need to be divided into task-based benchmarks to allow comparison with existing SLR systems.
Simultaneously, effects from real-world deployment and long-term usage feedback are currently not available and are required for a complete assessment of our contribution's practical utility.

\section{Discussion and Future Work\label{sec:future}}
The evaluation of SWARM-SLR and SLR automation in general remains difficult to scale.
It is either too time intensive if completed, or oversimplified and lacking context when segmented.
The SWARM-SLR AIssistant implementation is particularly suited for agentic or API-based tools, such as ORKG ASK or Zotero, has limited support for Python package wrapped solutions, such as SWARM-SLR's bnw-tools, and requires future work towards integrating more heterogeneous software, such as Miro or Obsidian.
The AIssistant is envisioned to be a uniform framework supporting any scholarly workflow.
Yet, severe limitations arise in the multi-objective optimization between energy/time/financial costs, efficiency, transparency, interoperability and reusability, clearness and many other factors mentioned in the UEQ and beyond.

\section{Conclusion\label{sec:conclusion}}
We presented the SWARM-SLR AIssistant, a modular agent-based framework integrating the SWARM-SLR workflow guidance.
Using a persistent data layer, we demonstrated the feasibility of aligning conversational LLM frameworks with the systematic literature review requirements of stage I.
Additionally, we proposed a tool registry to simplify tool annotation, enable decentralized contribution, and lower the threshold for extending research workflows.
18 respondents contributed to our preliminary evaluation, highlighting improved usability and accessibility, as well as challenges remain regarding tool transparency and resource efficiency.
While the SWARM-SLR AIssistant is a step towards accessible, modular SLR automation, achieving this goal will require significant future work towards a crowdsourced research software ecosystem facilitating accessible, yet transparent workflows.

\section*{Acknowledgment}
During the preparation of this work, the author(s) used \textbf{GPT-4.1 (GitHub Copilot)}, \textbf{DeepL}, \textbf{Grammarly (Browser Plugin)}, \textbf{LanguageTool (Browser Plugin)} in order to: \textbf{translate text}, \textbf{grammar and spelling check}, \textbf{paraphrase and reword}, according to the CEUR GenAI Usage Taxonomy\footnote{\url{https://ceur-ws.org/GenAI/Taxonomy.html}}.

\bibliographystyle{ieeetr}
\bibliography{main}

\end{document}